\documentclass[epjCONF]{svjour}
\usepackage{graphics}
\usepackage[varg]{txfonts} 
\usepackage[latin1]{inputenc}
\session-title{Conference Title, to be filled}
\begin{document}
\title{TeV gamma-UHECR  anisotropy by decaying nuclei in flight: \\
first neutrino traces? }
\author{Daniele Fargion\inst{1}\fnmsep\thanks{\email{daniele.fargion@roma1.infn.it}} }
\institute{Physics Department, Rome University 1, Sapienza,\and INFN, Roma 1}
\abstract{
Ultra High Cosmic Rays)  made by He-like lightest nuclei might  solve the AUGER extragalactic clustering along Cen A. Moreover He like UHECR nuclei cannot arrive from Virgo  because the light nuclei fragility and opacity above a few Mpc, explaining the Virgo UHECR absence.  UHECR signals are  spreading along Cen-A as observed because horizontal galactic arms magnetic fields, bending them on vertical angles. Cen A events by He-like nuclei are deflected as much as the observed clustered ones; proton will be more collimated while heavy (iron) nuclei are too much dispersed.  Such a light nuclei UHECR component coexist with the other Auger heavy nuclei and with the Hires nucleon composition. We foresaw (2009) that UHECR He from Cen-A AGN being fragile should partially fragment into secondaries  at  tens EeV  multiplet (D,$He^{3}$,p) nearly as the recent twin multiplet discovered ones (AUGER-ICRC-2011), at $20$ EeV  along Cen A UHECR clustering. Their narrow crowding  occur by a posteriori very low probability, below $3\cdot 10^{-5}$. Remaining UHECR spread group  may hint for correlations with other gamma  (MeV-$Al^{26}$ radioactive) maps, mainly  due to galactic SNR sources as Vela pulsar, the brightest, nearest GeV source. Other nearest galactic gamma sources   show links with UHECR via TeV correlated maps. We suggest that UHECR are also heavy radioactive galactic nuclei  as $Ni^{56}$,  $Ni^{57}$  and $Co^{57}$,$Co^{60}$ widely bent (tens degree up to $\geq 100^{o}$)  by galactic fields.  UHECR radioactivity (in $\beta$ and $\gamma$ channels) and decay in flight at hundreds keV  is boosted (by huge Lorentz factor $\Gamma_{Ni}\simeq 10^{9}- 10^{8}$)  leading to PeVs electrons and consequent synchrotron TeVs gamma offering UHECR-TeV correlated sky anisotropy. Moreover also rarest and non-atmospheric $\tau$, and e neutrinos secondaries at PeVs, as the first two rarest shower  just discovered in ICECUBE, maybe the first signature of such expected radioactive secondary tail.} 
\maketitle

\section{UHECR composition and maps}
UHECR astronomy is becoming a reality, suffering however by directionality  smearing of magnetic field along the UHECR arrival directions \cite{Auger-Nov07}.
   The UHECR compositions are leading to different bending and maps, different fragment and secondary gamma, neutrino spectra; these different nature are making UHECR nucleon origination
   well directed (but in a GZK bounded \cite{Greisen:1966jv} , tens Mpc distances, Universe) or even much local and smeared (a few Mpc) for our lightest UHECR nuclei. If UHECR are mostly heavy as Fe or Ni,Co, as argued  in present paper  then UHECR astronomy will be  so much bent and polluted to be greatly smeared  and hardly correlated to their sources. However very local galactic sources maybe recognized (see also \cite{Fargion2011c}). On the other side lightest nuclei UHECR astronomy may lead to a parasite  astronomy (i.e an astronomy associated due to secondaries fragments of UHECR as pions and consequent gamma, charged leptons and neutrinos) observable as a MeV-TeV gamma, UHE neutrino, and UHECR lightest nuclei fragment; these traces might be partially source of  radio, X  tails. This is the case of the nearest  extragalactic AGN, Cen A.  Extragalactic heavy nuclei may also be traced by nuclei fragments but at much lower energy. However we are considering also galactic UHECR heavy nuclei whose radioactivity is source of by beta decay electrons to synchrotron parasite gamma TeVs anisotropy in the sky.   Indeed we  address here on the UHECR anisotropy nature able  to  correlate UHECR maps and composition. Heavy radioactive nuclei  offer a reasonable tuned solution for most UHECR, excluding Cen A area. Among the boosted beta decay secondaries we foresee also a PeVs neutrino component. Its remarkable feature by mixing and oscillation is the presence of electron and tau flavor at PeV energy, that, contrary to muon ones, cannot be of terrestrial atmospheric nature. Two of these showering events maybe just the very first ones being found in last few weeks by ICECUBE \cite{ICECUBE-2012}.
\subsection{Smeared UHECR from Cen A by He and its fragments}
  Extragalactic UHECR from Cen A formed (mostly) by  lightest nuclei may explain a partial clustering of events, as the one around CenA \cite{Auger10},\cite{Fargion2011},  as well as the puzzling UHECR absence  around Virgo \cite{Auger-Nov07}. Light nuclei are fragile and fly few Mpc before being    halted by photo-disruption \cite{Fargion2008}.   Of course a more heavy UHECR nature may lead to a complete  or partial confinement explaining in alternative the Virgo absence.  The light nuclei fragments $He + \gamma \rightarrow D+D,D+\gamma \rightarrow p+n+ \gamma  , He + \gamma \rightarrow He^{3}+n, He + \gamma \rightarrow T +p $  may nevertheless trace the same UHECR maps by a secondary  clustering at half or even fourth of the UHECR primary  energy. Moreover also a rare neutron (ten EeV) decay in flight (hundred kpc decay distance) component $He + \rightarrow He^{3}+n, n \rightarrow p+e+ \bar{\nu_{e}}$ may shine by tens PeV electrons via synchrotron radiation, into TeVs gamma anisotropy, as the recent ICECUBE TeV anisotropy show.  At lower energy (at ten EeV or below) the huge smeared cosmic ray isotropy and homogeneity may hide these tiny inhomogeneity traces  \cite{Fargion2010},\cite{Fargion2011}. About the extragalactic hypothesis let us remind that gamma UHECR secondaries rays  are partially absorbed by microwave and infrared background making once again a very local limited UHECR-gamma astronomy. Muon neutrinos $\nu_{\mu}$, the most penetrating and easy to detect on Earth, are unfortunately deeply polluted by atmospheric (homogeneous) component (as smeared and as isotropic as their parent charged CR nucleons and nuclei ). This atmospheric neutrino isotropy and homogeneity  (made by primary CR bent and smeared by galactic magnetic fields)  is well probed by last TeV muon neutrino smooth ICECUBE map \cite{ICECUBE-11}. Tau neutrinos on the contrary, the last neutral lepton discovered, are almost absent in atmospheric secondaries. At a few GeV energies or below $\tau$ cannot be born because energy thresholds. Around tens GeV atmospheric windows, $\nu_{\mu}\rightarrow \nu_{\tau} $ neutrino oscillation   may  arise because the Earth size is large  enough to allow a complete neutrino $\nu_{\mu}\rightarrow \nu_{\tau} $  conversion; however at higher energy, tens TeV-PeV up to PeV ore EeV  $\nu_{\tau} $ atmospheric neutrinos cannot convert within short Earth size (from $\nu_{\mu}\rightarrow$ $\nu_{\tau}$) because they are unable to oscillate at such high energies; therefore $\nu_{\tau}$ neutrino might be a clean signal of  UHECR-neutrino associated astronomy\cite{FarTau}.
The tau birth in ice or sea  may shine as a double bangs (disentangled above PeV)  \cite{Learned} or in a \emph{mini twin bangs} observable in Deep Core  or PINGU detector \cite{Fargion2011c}, or ANTARES \cite{antares}. In addition high energy UHE $\nu_{\tau}$ and its ${\tau}$, born tangent to the  Earth or mountain, while escaping  in air  may lead, by decay in flight, to  loud, amplified  and well detectable directional tau-airshower at horizons \cite{Fargion1999},\cite{FarTau} detectable by Cherenkov lights as in ASHRA experiment \cite{Aita11} \cite{FarTau}, or by  ground detectors and fluorescence lights \cite{FarTau},\cite{Bertou2002} \cite{Feng02}. Tau astronomy versus UHECR are going to reveal most violent  sky as the most deepest probe. This tau airshowers or Earth skimming neutrino were considered since more than a decade and are going to be observed in AUGER \cite{Auger08} or TA \cite{TA-12} in a few years \cite{Fargion1999},\cite{FarTau},\cite{Auger-01},\cite{Feng02},\cite{Auger08}.
Finally PeV neutrino showers in cubic km if of muon flavor, might be atmospheric or  prompt neutrino, while if they are not tracing any muon tail, they maybe more probably extraterrestrial, either electron or tau in nature. This maybe just the case of last two discovered ones by ICECUBE few weeks ago.
\subsection{ Gamma and UHECR maps. may them correlate?}
     We  found that Cen A (the most active and nearby extragalactic AGN) is apparently the most shining  UHECR source whose clustering (almost a quarter of the event) along a narrow solid angle  (whose opening angular size is  $\simeq 17^{o}$) \cite{Auger10}, is convincing and in agreement with lightest nuclei \cite{Fargion2008}, \cite{Fargion09b}.
     However the main question is related to remaining majority of events. Where do they come from?
          In recent maps of UHECR we noted a first hint of  Vela  where the brightest and  the nearest gamma source, is associated to an unique UHECR triplet nearby the pulsar \cite{Fargion09b}. The correlation is also based on MeV  Comptel map   coincidence. The needed UHECR  bending from near Vela ($815$ yc far away) is requiring a very heavy nuclei (or light nuclei with large magnetic field) discussed below.   Remaining UHECR events might be also mostly   heavier nuclei more bent and smeared by galactic fields. To exhibit some clustering (as some where they show)  they need to be mostly galactic. Let us remind that UHECR events initially consistent with  GZK volumes  \cite{Auger-Nov07}, today seem to be not much correlated with expected Super Galactic Plane \cite{Auger10}. Moreover slant depth data of UHECR from AUGER airshower shape do not favor the proton but point to  a nuclei. Therefore in the present paper we suggest UHECR as made by nuclei, light from Cen A and heavier (Ni,Co) mostly from our galaxy. It maybe worth to briefly remind the long (two decades at least) history of the UHECR understanding (miss-understanding)  somehow linked to a century old (Victor Hess first balloon detection) cosmic ray puzzle, but it will take us too far away from present short contribute.

\section{The UHECR galactic bending for $He^{4}$ and  $Ni^{57}$ }
Cosmic Rays are blurred by magnetic fields. Also UHECR suffer of a Lorentz force deviation. This smearing maybe source of UHECR features. Mostly along Cen A. There are at least three mechanisms for magnetic deflection along the galactic plane, a sort of galactic spectroscopy of UHECR \cite{Fargion2008}. The magnetic  bending by extra-galactic fields are  in general negligible respect galactic ones.  A late nearby (almost local) bending by a nearest coherent galactic arm field, and a random one by turbulence and a random along the whole plane inside different arms:\\
(1) the coherent Lorentz angle bending $\delta_{Coh} $ of a proton (or nuclei) UHECR (above GZK \cite{Greisen:1966jv}) within a galactic magnetic field  in a final nearby coherent length  of $l_c = 1\cdot kpc$ is: \\$ \delta_{Coh-p} .\simeq{2.3^\circ}\cdot \frac{Z}{Z_{H}} \cdot (\frac{6\cdot10^{19}eV}{E_{CR}})(\frac{B}{3\cdot \mu G}){\frac{l_c}{kpc}}$\\
(2) the random bending by random turbulent magnetic fields, whose coherent sizes (tens parsecs) are short and whose final deflection angle is  smaller than others and they are here ignored;\\
(3) the ordered multiple UHECR bending along the galactic plane across and by alternate arm magnetic field directions whose final random deflection angle is remarkable and discussed below.\\The bending angle value is  quite different for a heavy nuclei as an UHECR from Vela whose distance is only $0.29$ kpc:
 $\delta_{Coh-Ni} \simeq
{18,7^\circ}\cdot \frac{Z}{Z_{Ni^{28}}} \cdot (\frac{6\cdot10^{19}eV}{E_{CR}})(\frac{B}{3\cdot \mu G})({\frac{l_c}{0.29 kpc})}$\\
  Note that this spread is  able to explain the nearby Vela TeV anisotropy (because radioactive emission in flight) area around its correlated UHECR triplet.There is an extreme possibility: that Crab pulsar at few kpc is feeding the TeV anisotropy connecting with a gate its centered disk to a wider extended region where some UHECR are clustering. From far Crab distances the galactic bending is:  $\delta_{Coh-Ni} \simeq {129^\circ}\cdot \frac{Z}{Z_{Ni^{28}}} \cdot (\frac{6\cdot10^{19}eV}{E_{CR}})(\frac{B}{3\cdot \mu G})({\frac{l_c}{2 kpc})}$\\
  Note that such a spread is able to explain the localized TeV anisotropy born in Crab (2 kpc) apparently  extending  around  area near Orion, where also spread UHECR events seem clustered.Such heavy iron-like  (Ni,Co) UHECR , because of the big charge and large angle bending, are mostly bounded inside a Galaxy, as well as in Virgo cluster, possibly explaining the UHECR  absence in that direction.
 The incoherent random angle bending (2) along the galactic plane and arms, $\delta_{rm} $, while crossing along the whole Galactic disk $ L\simeq{20 kpc}$  in different (alternating) spiral arm fields   and within a characteristic  coherent length  $ l_c \simeq{2 kpc}$ for He nuclei is $$\delta_{rm-He} \simeq{16^\circ}\cdot \frac{Z}{Z_{He^2}} \cdot (\frac{6\cdot10^{19}eV}{E_{CR}})(\frac{B}{3\cdot \mu G})\sqrt{\frac{L}{20 kpc}} \sqrt{\frac{l_c}{2 kpc}}$$ The heavier  (but still light nuclei) bounded from Virgo might be also Li and Be:
 $\delta_{rm-Be} \simeq{32^\circ}\cdot \frac{Z}{Z_{Be^4}} \cdot (\frac{6\cdot10^{19}eV}{E_{CR}})(\frac{B}{3\cdot \mu G})\sqrt{\frac{L}{20 kpc}}
\sqrt{\frac{l_c}{2 kpc}}$.  It should be noted that the present anisotropy above GZK \cite{Greisen:1966jv} energy $5.5 \cdot 10^{19} eV$ (if extragalactic)  might leave a tail of signals: indeed the photo disruption of He into deuterium, Tritium, $He^3$ and protons (and unstable neutrons), rising as clustered events at half or a fourth (for the last most stable proton fragment) of the energy:\emph{ protons being with a fourth an energy but half a charge He parent may form a tail  smeared around Cen-A at twice larger angle} \cite{Fargion2011}. We suggested  to look for correlated tails of events, possibly in  strings at low $\simeq 1.5-3 \cdot 10^{19} eV$ along the Cen A train of events. \emph{It should be noticed that Deuterium fragments have half energy and mass of Helium: Therefore D and He spot are bent at same way and overlap into UHECR circle clusters} \cite{Fargion2011}.  Deuterium are even more bounded in a very local Universe because their fragility (explaining Virgo absence). In conclusion He like UHECR may be bent by a characteristic angle as large as $\delta_{rm-He}  \simeq 16^\circ$; its expected lower energy Deuterium or proton fragments at half energy ($30-25 EeV$) are also deflected accordingly at ($\delta_{rm-p}  \simeq 16^\circ$); protons last traces at a quarter of the UHECR energy, around twenty EeV energy, will be spread within ($\delta_{rm-p}  \simeq 32^\circ$), within the observed Cen A UHECR multiplet shown in figures, see Fig. \ref{figure1}.
 \subsection{Twin UHECR multiplet at 20 EeV pointing Cen A}
\begin{figure}\sidecaption
\resizebox{0.45\hsize}{!}{\includegraphics*{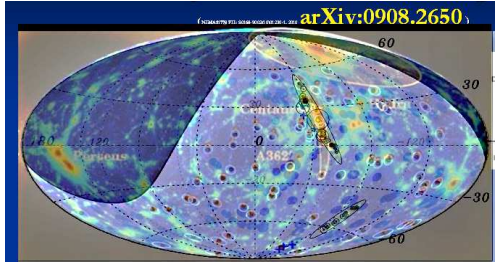}}
\caption{The last 2010 UHECR event map \cite{Auger10}, white rings,  and the multiplet event map \cite{Auger11} over the infrared background and the UHECR tens EeV multiplet. It is remarkable the absence of the expected wide Virgo area (on the top). The observed twin multiplet at $20$ EeV lay over an earlier predicted twin one (banana-like) slightly different, foreseen in  earlier articles.\cite{Fargion2009-2010}. Note also the third multiplet along Large and Small Magellanic Clouds, is suggesting also a galactic UHECR component. Finally note the presence of a unique AUGER-TA doublet along the galactic plane: both exceeding $10^{20}$ eV, making a galactic source as Aq X1, a source candidate\cite{Troitsky}}\label{figure1}
\end{figure}

 The very  recent multiplet clustering published just few weeks ago by AUGER at twenty EeV  contains just three and apparently isolated train of events  with  (for the AUGER collaboration) no statistical meaning. \cite{Auger11}. Indeed apparently they are pointing to unknown sources (See Fig. \ref{figure1}). However the crowding of the two train multiplet tail centers inside a very narrow disk area focalized about the rarest Cen A UHECR source is remarkable, \cite{Fargion2011}.  If UHECR are  made by proton (as some AUGER author believe) they will not naturally explain such a tail structure because these events do not cluster more than a few degree, contrary to observed UHECR and associated multiplet (See Fig. \ref{figure1}). Also heavy nuclei whose smearing is much larger and whose eventual nucleon fragments ($A\rightarrow (A-1)$) should lead to parasite tail that greatly differs in mass and energy and bending angle with the observed AUGER one. If heavy UHECR are around us and they are bent, only a galactic smeared component may be somehow discovered. Our He-like UHECR do fit the AUGER and the HIRES composition traces. The He secondaries are splitting in two (or a fourth) energy fragments along Cen A tail   whose presence has  being foreseen and published many times  in last years\cite{Fargion2009-2010},\cite{Fargion2011}. Indeed the  dotted circle around Cen A containing the two (of three) multiplet  has a radius as small as $7.5^{o}$,  it extend in an area that is as smaller as  $200$ square degrees, below or near $1\% $ of the observation AUGER sky. The probability that two among three sources fall inside this small area is offered by the binomial distribution. $$ P (3,2) = \frac{3!}{2!} \cdot (10^{-2})^{2} \cdot \frac{99}{100}\simeq 3 \cdot 10^{-4}$$ Moreover the same twin tail of the events are aligned almost exactly $\pm 0.1 $ rad along UHECR train of events toward Cen A. Therefore the UHECR  multiplet alignment  at twenty EeV has an a priori  probability as low as $ P (3,2) \simeq 3 \cdot 10^{-5}$ to follow the foreseen signature\cite{Fargion2011}.

\begin{figure}\sidecaption
\resizebox{0.45\hsize}{!}{\includegraphics*{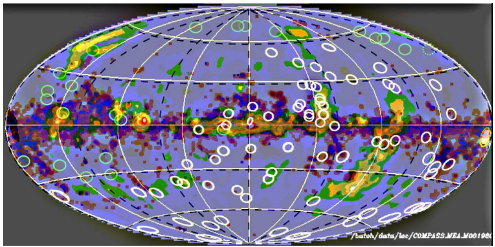}}
\caption{The last 2010 UHECR event map \cite{Auger10}  over the last Planck cold cloud map and the Comptel Gamma map; also the most recent Telescope Array, TA, UHECR events are added in the North celestial sky \cite{TA-2012}. There is a probable  correlation with the enhanced gamma regions and UHECR events. \cite{Fargion2011}. Note also the gamma trace along the Magellanic stream linking Large and Small Clouds.}\label{figure2}
\end{figure}

\begin{figure}\sidecaption
\resizebox{0.7\hsize}{!}{\includegraphics*{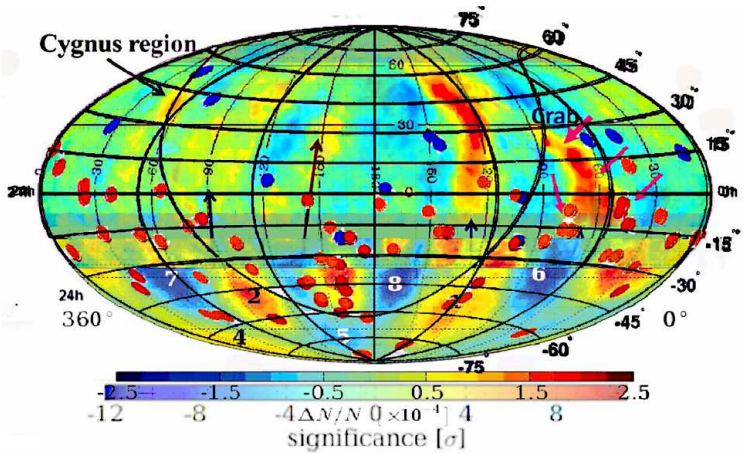}}
\caption{The last 2010 UHECR event (red disks) map \cite{Auger10} and Hires (blue) ones over the TeV sky anisotropy as observed in the North by ARGO and in the South by ICECUBE atmospheric muon traces, in celestial coordinate. The TeV-UHECR connection is obvious}\label{figure3}
\end{figure}

\section{TeV Gamma and UHECR  nuclei connection? }
       In recent maps of UHECR we noted first hint of galactic source   rising as an UHECR triplet \cite{Fargion09b}.  Also the hint by $Al^{26}$ gamma map traced by Comptel somehow overlapping with UHECR events at 1-3 MeV favors a role of UHECR radioactive elements (as $Al^{26}$), see Fig. \ref{figure2}.   The most prompt ones are the  $Ni^{56}$, $Ni^{57}$ (and $Co^{56}$, $Co^{60}$ ) made by Supernova (and possibly by their collimated GRB micro-jet components, see \cite{Fargion1998}) ejecta in our own galaxy.  Indeed in all SN Ia models, the decay chain  $Ni^{56}\rightarrow Co^{56} \rightarrow Fe^{56}$provides the primary source of energy that powers the supernova
optical display. The $Ni^{56}$ decays by electron capture and the daughter $Co^{56}$  emits gamma rays by the nuclear de-excitation
process; the two characteristic gamma lines are  respectively at $E_{\gamma} =$ 158 keV and $E_{\gamma}=$ 812 keV. Their half lifetime are spread from $35.6$ h for  $Ni^{57}$ and $6.07$ d. for   $Ni^{56}$. However there are also more unstable radioactive rates as  for $Ni^{55}$ nuclei whose half life is just $0.212$ s or $Ni^{67}$ whose decay is $21$ s. Therefore we may have an apparent boosted  ($\Gamma_{Ni^{56}} \simeq 10^{9}$) life time spread from $2.12 \cdot 10^{8}$ s or $6.7$ years (for $Ni^{55}$) up to nearly  $670$ years (for $Ni^{67}$) or $4$ million years for  $Ni^{57}$. This consequent wide range of lifetimes guarantees a long life activity on the UHECR radioactive traces. However the most bright are the fastest  decaying ones. The arrival tracks of these UHECR radioactive heavy nuclei may be  widely bent, as shown below, by galactic magnetic fields. Among the excited nuclei to mention  for the UHECR-TeV connection is $Co_{m}^{60}$  whose half life is $10.1$ min and whose decay gamma line is at $59$ keV. At a boosted nominal Lorentz factor $\Gamma_{Co^{60}}= 10^{9}$ we obtain $E_{\gamma}\simeq 59 $ TeV ; let us remind that a gamma air-shower exhibit a smaller secondary muon abundance with respect to an equivalent hadronic one; therefore a gamma  simulates a ($10\%$) hadronic shower ( $E_{gamma-hadron}\simeq 6 $ TeV) nearly corresponding to observed ICECUBE-ARGO anisotropy \cite{Desiati},\cite{ARGO}. The decay boosted lifetime is $19000$ years corresponding to  6 kpc distance. Therefore $Co_{m}^{60}$ energy decay traces, lifetime and spectra fit well within  present UHECR-TeV connection for nearby galactic sources as Vela and (probable) Crab. Other radioactive beta decay, usually at higher energy may also shine at hundred or tens TeV or below by inverse Compton and synchrotron radiation. Therefore their UHECR bent parental nuclei may shine also in TeV Cosmic ray signals.


\section{Conclusions: UHECR-TeV by ultra-relativistic $Ni^{57}$ decay? }
The very rich UHECR map of AUGER and HIRES in celestial coordinate overlapped on TeV anisotropy  (North sky by ARGO- South sky by ICECUBE atmospheric muons) is displayed in next figure, see Fig.\ref{figure3}.
 The figure shows a clear area around Crab, marked by arrows, that is somehow extending in a wide anisotropy area where few UHECR took place. We also added arrows to  remind the Cen A unique clustering as well as the Vela and the Cygnus galactic TeV sources, somehow in connection with UHECR.


  Surprisingly the UHECR puzzle maybe at a corner stone: the UHECR-Multiplet along Cen A, the absence of Virgo, the hint of correlation with  Vela and  with galactic TeV anisotropy \cite{Desiati} \cite{ARGO}, might be in part solved by an extragalactic lightest nuclei, mainly He, from Cen A ,A partial confirm is the predicted \cite{Fargion2011} and observed \cite{Auger11} multiplet clustering by fragments (D,p) at half UHECR edge energy aligned   with Cen A: He like UHECR  may be bent by a characteristic angle as large as  $\delta_{rm-He}  \simeq 16^\circ$ \cite{Fargion2011} while their fragments multiplet  follow along a tail spread by a wider angle $\delta_{rm-p}  \simeq 32^\circ$ \cite{Fargion2011},\cite{Auger11}. also neutron beta decay may feed a TeV correlated anisotropy.  Other UHECR spread events, might be due to a  dominant heavy radioactive nuclei component  $Ni^{56}$, $Ni^{57}$ and $Co^{56}$, $Co^{60}$,originated  by galactic sources  (old SNR-GRB relics) as suggested also by relic $Al$ nuclei at rest in gamma map. UHECR Ni,Co maybe deflected by $18,7^{o}$ for Vela,  $128^{o}$ (or less) for Crab tuning within TeV inhomogeneities, made by boosted hundred keV gamma and beta positrons decay, shining at TeVs .  We predict here analogous UHECR traces around Cygnus and Cas A in future TA UHECR  maps. Inner galactic core UHECR are widely spread and hidden by magnetic fields in dense magnetic galactic core arms. However more clustering around ($\geq 20^{o}$) the galactic plane far from the core, is expected in future data. Magellanic cloud and stream may rise in UHECR maps. UHECR should rise around Cas A and Cygnus, seen by T.A. in North sky. Recent doublet toward  Aq X1 may be a new galactic source.
   The UHECR spectra cut off maybe not indebt to the expected extragalactic GZK feature but to the more modest imprint of a galactic confinement and of nuclei spectrography.    The UHECR radioactive beta decay in flight may trace in new $\nu_{\tau}$ neutrino astronomy or anisotropy, noise free,  related to astronomical (parasite oscillated) tau neutrino;   boosted tau (\emph{mini-double bang} \cite{Learned}, within a 5 meter size) in Deep Core  or ANTARES \cite{antares} may reveal hundred TeV tau decay (seeing  similar PeV ones in ICECUBE \cite{Learned}). Also Tau airshowers may rise in Cherenkov beamed air-showers. \cite{Fargion1999}, \cite{FarTau}, as being searched in ASHRA experiment, \cite{Aita11} or in fluorescence telescopes for higher tau energies \cite{FarTau},\cite{Feng02},\cite{Bertou2002},\cite{Auger07}, \cite{Auger08}.  The discover of such expected  Neutrino astronomy  may shed additional light on the UHECR nature, origination  and mass composition, while  opening our eyes to mysterious  UHECR sources. Future gamma data and UHECR correlation, additional multiplet  may  lead to a more conclusive fit of this unsolved, century old, cosmic ray puzzle. Tau astronomy (observable also by mini-twin double bangs in Deep Core or Tau Airshowers) as well as ICECUBE showering at PeV with no muon may also offer an additional windows for the first extraterrestrial neutrino traces \cite{ICECUBE-2012}.\\

\clearpage

\end{document}